\documentclass[%
 reprint,
 amsmath,amssymb,
 aps,
]{revtex4-2}

\usepackage{graphicx}% Include figure files
\usepackage{dcolumn}% Align table columns on decimal point
\usepackage{bm}% bold math
\begin{document}

\preprint{APS/123-QED}

\title{Investigate the Performance of Distribution Loading with Conditional Quantum Generative Adversarial Network Algorithm on Quantum Hardware with Error Suppression}% Force line breaks with \\

\author{Anh Pham}
 \affiliation{Deloitte Consulting LLP}
 
\author{Andrew Vlasic}%
\affiliation{Deloitte Consulting LLP}

\date{\today}% It is always \today, today,

\begin{abstract}
The study examines the efficacy of the Fire Opal error suppression and AI circuit optimization system integrated with IBM's quantum computing platform for a multi-modal distribution loading algorithm. Using Kullback-Leibler (KL) divergence as a quantitative error analysis, the results indicate that Fire Opal can improve on the time-dependent distributions generated by our Conditional Quantum Generative Adversarial algorithm  by 30-40\% in comparison with the results on the simulator. In addition, Fire Opal's performance remains consistent for complex circuits despite the needs to run more trials. The research concludes that Fire Opal's error suppression and circuit optimization significantly enhanced quantum computing processes, highlighting its potential for practical applications. In addition, the study also reviews leading error mitigation strategies, including zero noise extrapolation (ZNE), probabilistic error cancellation (PEC), Pauli twirling, measurement error mitigation, and machine learning methods, assessing their advantages and disadvantages in terms of technical implementation, quantum resources, and scalability.

\end{abstract}

\maketitle

%\tableofcontents

\section{Introduction}
Data loading is critical for many quantum algorithms and applications. However, it is a challenging problem when executing on NISQ quantum hardware due to the longer circuit depth in this quantum subroutine. In this report, we demonstrate the utility of error suppression and AI-enabled circuit optimization in Fire Opal to load multi-modal distributions on IBM Kyoto as implemented in our Conditional Quantum Generative Adversarial Network (C-QGAN) algorithm. Specifically, the distributions generated with Fire Opal on quantum hardware produced results much closer to the ideal distributions, and showed an approximate $30\%-40\%$ improvement in results relative to running without it based KL divergence analysis.

Previously we proposed a new quantum algorithm known as C-QGAN \cite{certo2023conditional} to load multiple non-uniform distributions using condition registers. This data loading technique was then applied in conjunction with quantum amplitude estimation \cite{montanaro} to evaluate complex financial instrument called Asian option. The state preparation based on C-QGAN was shown to be less computationally expensive than other well known technique like Grover-Rudolph \cite{grover2002creating} and QGAN \cite{mirza2014conditional} which can potentially negate the quadratic speedup of algorithm like QAE \cite{herbert2021problem} when they are approximating stochastic processes.

Due to the noisy nature of current NISQ hardware, various error mitigation techniques have been applied to enable improve the outputs when quantum algorithms are run on quantum hardware. Error mitigation techniques are post-processing algorithms, which happen at the software level, that can improve on the distorted values obtained from quantum hardware due to different noise sources. In addition, these techniques have been applied to improve  in many variational quantum algorithms \cite{bhattacharjee2024enhancing} which have many important applications in quantum machine learning, chemistry and optimization. However, a potential drawbacks of many error mitigation techniques can be that they require extra classical and quantum computational resources to recover the noise-free values, which limits their applications only for short-depth circuits \cite{RevModPhys.95.045005}. As a result, our report aims to explore error suppression technique, which happens at the hardware level, as implemented in Fire Opal to understand their utility in the state-preparation process. 

%We have tested the performance of the error suppression system Fire Opal on IBM’s quantum cloud. Fire Opal was very easy to use, demonstrated an approximate $30\%-40\%$ improvement in results relative to running without it, and produced results much closer to the ideal distributions.

%To test the performance of Fire Opal on IBM’s quantum cloud the process known as ``distribution loading” was ran with and without Fire Opal, and with a moderate depth circuit and deep depth circuit. Particularly, we trained four time-dependent distributions through a conditional quantum generative adversarial network (C-QGAN). Once mapped onto the physical hardware in a process called transpilation, the complexity of the circuits increased significantly.

%We found that Fire Opal’s integration with the IBM cloud system is straightforward and does not require user input. It optimizes the logical circuit and automatically applies error suppression during execution, optimizing the quantum computing calculations. Furthermore, quantitative error analysis using KL divergence showed that the distribution generated with Fire Opal was significantly closer to the results from an ideal, noise-free environment. The results demonstrated an improvement of approximately $30\%-40\%$ when compared to results generated from the IBM system without Fire Opal on all four distributions.

\section{Overview of Error Mitigation Methods}

\begin{figure*}[ht]
    \centering
    \includegraphics[width=1.0\textwidth]{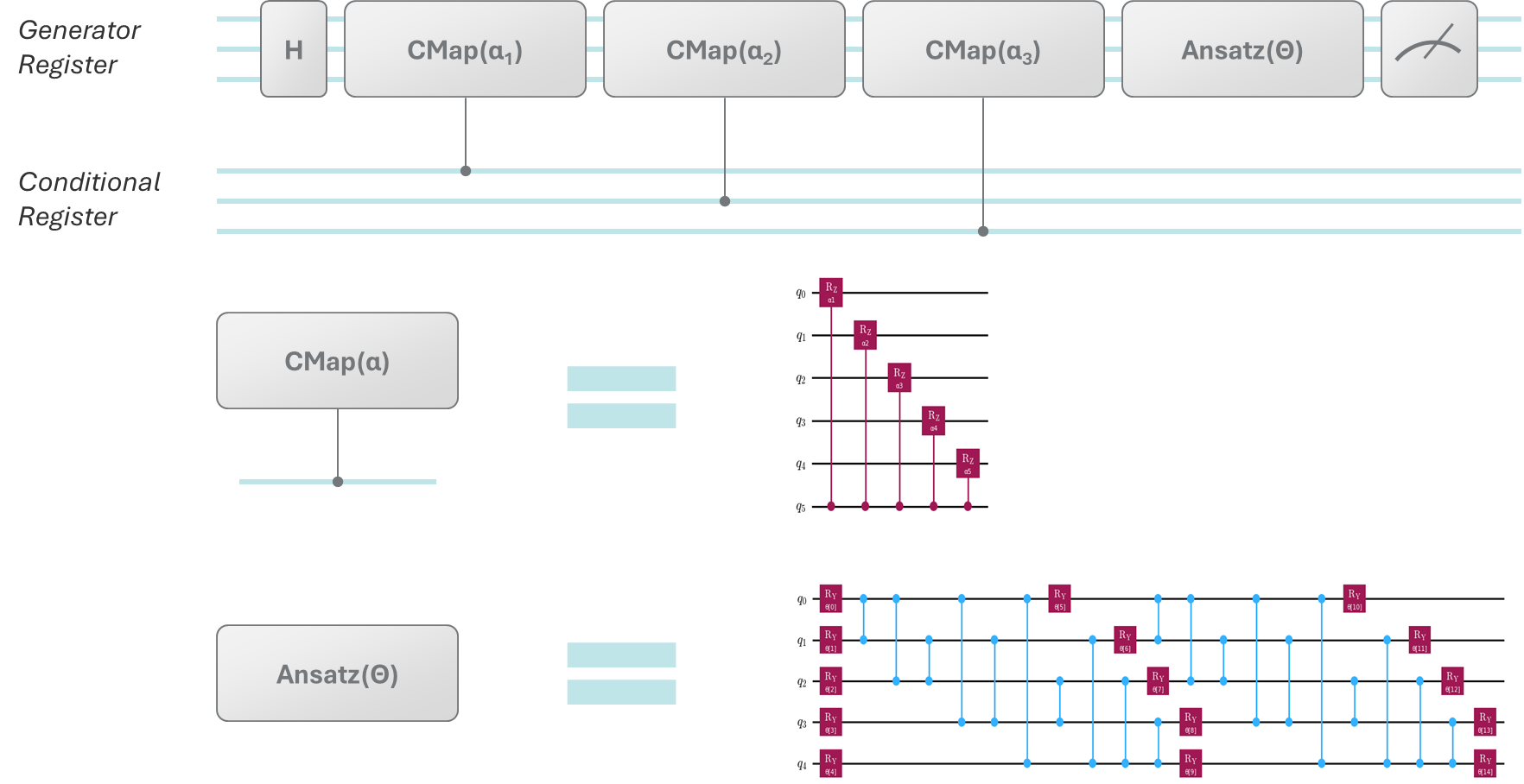}
    \caption{\label{fig:circ} Circuit to test the technique in Certo et al. \cite{certo2023conditional}. Here the circuit depth of the ansatz Two-Local circuit was doubled, resulting in three layers of $Ry$ gates and two la
    yer of one-to-all control-$Z$ gates in-between the $Ry$ layers}
\end{figure*}

For completeness, we list and describe different class of error mitigation. Currently, the established methods for error mitigation lie in one of the five classes: 
\begin{enumerate}
    \item Zero Noise Extrapolation (ZNE);
    \item Probabilistic Error Cancellation (PEC);
    \item Pauli Twirling;
    \item Measurement Error Mitigation; and,
    \item Machine Learning techniques.
\end{enumerate}

Each of the error mitigation classes have their respective advantages and disadvantages, ranging from ease of implementation and minimal number of assumptions, and increased time to run an algorithm and exponential increase of gates to circuits. Below the advantage and shortcomings of each class is given. 

\paragraph{ZNE:} ZNE works with unknown noise models, and it can be used with different algorithms, including variational methods. In addition, it can be applied through digital or analog modes. However, assuming the noise model is time invariant, ZNE needs to run the experiment several times. Moreover, ZNE can be sensitive to the methods used to amplify noise and needs to be used with other error mitigation methods since it won't be able to correct for error from state preparation and measurement (SPAM) \cite{giurgica2020digital,kandala2019error,bhattacharjee2024enhancing,majumdar2023best}.

\paragraph{PEC:} PEC can accurately correct for noise for an arbitrary number of qubits and is especially helpful for local dephasing noise and crosstalk errors. Given these advantages, PEC sampling overhead scales exponentially with error rates and circuit depth. Application may be limited to noise cancellation in simulating open quantum dynamics \cite{van2023probabilistic,ma2024limitations,mcdonough2022automated,takagi2022fundamental}. 

\paragraph{Pauli Twirling:} This technique is easy to implement since it involves converting different noise models (Markovian, non-Markovian, time dependent error) into Pauli noise channels to reduce coherent noise by modifying the circuits with appropriate Pauli twirls. In addition, it can be used for different applications and in conjunction with other error mitigation strategies. However, Pauli Twirling needs to be used with an efficient compiler to maintain circuit depth. Otherwise, the circuit depth will increase with the single-qubit Pauli operators. In addition, the implementation of Pauli Twirling requires executing multiple randomized circuits with random Pauli gates which can increase overhead cost. \cite{hashim2020randomized,chen2023learnability,cai2020mitigating,cai2019constructing,santos2024pseudo}.

\paragraph{Measurement Error Mitigation} A general assumption about the noise, with some restrictions; only the expectation is analyzed, making the technique purely classical; there are different implementations including the assumption of independence, modeling the assignment matrix with a continuous time Markov process, solving via constrained optimization, or using classical neural network to parse the data and separate the noise. But, overhead increases exponentially with the circuit size or number of qubits, has added steps increase time to solution, and some methods require a trained classical model \cite{kim2020quantum,geller2020rigorous,nachman2020unfolding,palmieri2020experimental,czarnik2021error,lowe2021unified,pokharel2024scalable}.

\paragraph{Machine Learning Techniques} The techniques leverage neural network architecture to mitigate the noise, using a minimal number of assumptions about the noise. Furthermore, pretrained networks allow for transfer learning to new circuits, considerably decreasing time to retrain the network. Autoencoders tend to play a prominent role in this technique. However, machine learning adds at least polynomial growth to the length of the circuit through the extra gates added to the circuit, and is a variational model that requires data and time to train. Finally, statistical models have a large potential to be undertrained and biased \cite{wang2022multidimensional,locher2023quantum,chalkiadakis2023quantum}.

\section{Methodology}

\begin{figure*}[ht]
    \centering
    \includegraphics[width=1.0\textwidth]{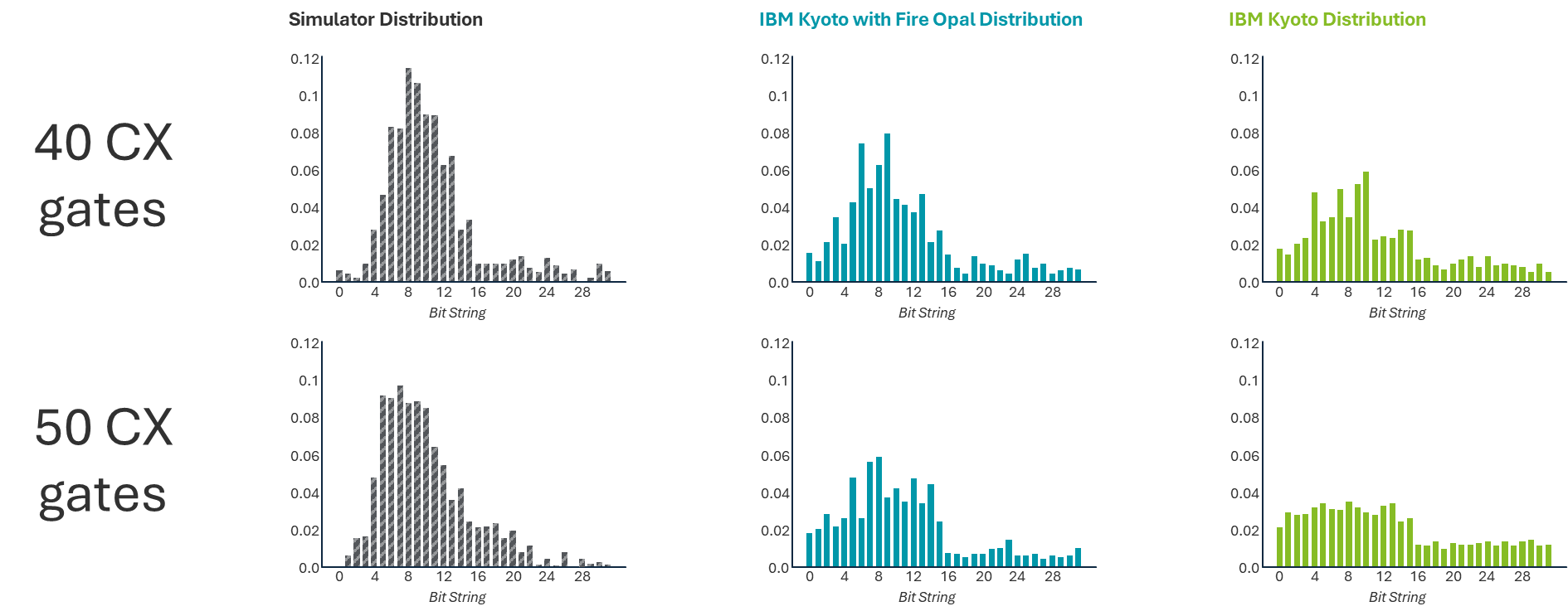}
    \caption{\label{fig:dist} Distribution from two different gate depths calculated from the simulator, on IBM Kyoto processor, and IBM Kyoto processor with Fire Opal.}
\end{figure*}

\begin{figure*}[ht]
    \centering
    \includegraphics[width=1.0\textwidth]{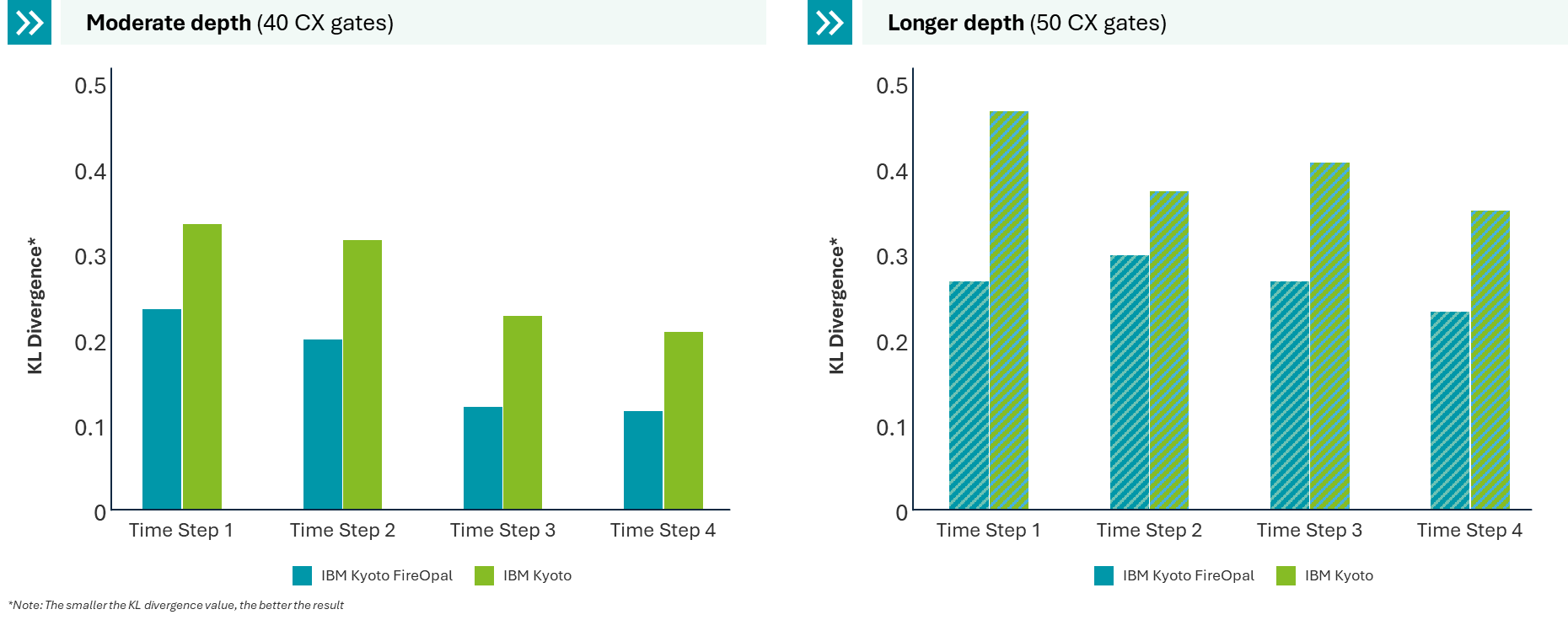}
    \caption{\label{fig:kl} KL divergence with all four distributions in the circuit calculated from the simulations against either the calculations from IBM Kyoto with Fire Opal and calculations From IBM Kyoto without Fire Opal.}
\end{figure*}

To understand how a blackbox model of error suppression and circuit optimization can be used to enhance the data loading process on NISQ hardware, we performed the inference based on our pretrained C-QGAN circuit on IBM Kyoto with and without Fire Opal. Before running these tests, we trained circuits using a quantum simulator to obtain optimal parameters and then tested on actual hardware. The results were then compared against the outputs generated on Qiskit statevector simulator, which theoretically generates perfect results without noise. The performance of Fire Opal was then quantified by using the Kullback-Leibler (KL) divergence to quantify the difference between the the simulator results and the results on quantum hardware with and without Fire Opal.

Four distributions corresponding to different time-steps were loaded with controlled registers corresponding to respective time steps given in \cite{certo2023conditional}; the circuit is displayed in Figure \ref{fig:circ}. Specifically, as shown in Fig. \ref{fig:circ}, the circuit is composed of generator registers (5 qubits) and controlled registers (2 qubits), whereby the generator anstaz is based on the Two-Local circuit in Qiskit. The Two-Local subprocess consists of two layers of $Ry$ gates and a layer of one-to-all control-$Z$ gates in-between the $Ry$ layers. In our experiments, we fully entangled all the qubits. Further explaination of the algorithm and circuit structures can be found in \cite{certo2023conditional}

We used two different circuits with increasing depths to understand the performance of Fire Opal with increasing number of two-qubit gates by increasing the depth of the ansatz layer in the generator. Specifically, two different circuit depths were tested. The moderate circuit depths contained a total number of gates at 120 (40 two-qubit gates), and much deeper circuit with 205 (50 two-qubit gates). In addition, the moderate circuit was sampled with 4096 shots, and the longer circuit was sampled with 8000 shots. 

As our baseline, we ran the C-QGAN circuits on IBM Kyoto without error suppression techniques. In order for the quantum circuits to be executed on quantum hardware, the original circuits need to be \textit{transpiled} to match IBM Kyoto's topology, and its native gates. To maximize the performance, we aimed to reduced the total number of gates, especially the two-qubit gates to reduce the noise contribution. As a result, we used the highest level of optimization possible, which is level  3. This resulted in a total number of native gates of 784 for the moderate depth circuit, and 997 for the longer circuit. In addition, since this is the baseline experiments, we did not include any error mitigation so that we can quantitatively asses the performance of Fire Opal. When the circuits were ran with Fire Opal, the original circuits were used as input without transpilation since Fire Opal has a function for circuit optimization. To activate Fire Opal the function on IBM cloud, the settings were changed on the instance to ``q\textunderscore ctrl”.

\section{Results and Discussions}
%As previously noted, the circuits were pretrained on a simulator. Inference was performed on IBM Kyoto with and without Fire Opal to compare with simulator results.

Analysis of testing results shows clear improvement in the Fire Opal distribution compared to IBM Kyoto by itself; see Figure \ref{fig:kl} for a visual of the analysis.  Fire Opal reduced noise and yielded better overall results on IBM Kyoto by around $30\%-40\%$. Specifically, the distributions generated on IBM Kyoto with Fire Opal qualitatively captures the specific shape of the distribution with peaks and the long tail. This becomes more pronounced as the number of two-qubit gates.  

In conclusion, our experiment validates the effectiveness of Fire Opal's error suppression and circuit optimization capabilities, highlighting potential to enhance the utilities of quantum hardware in the NISQ era.
%Performance was quantified by calculating the KL divergence between the distributions generated from the IBM hardware with and without Fire Opal, in comparison to the results on the simulator, for the two different circuit depths.

\pagebreak

\bibliography{apssamp}% Produces the bibliography via BibTeX.

\end{document}